\documentclass[aps,pre,twocolumn,showpacs,superscriptaddress,groupedaddress]{revtex4}

\newcommand{\beq}{\begin{eqnarray}}
\newcommand{\eeq}{\end{eqnarray}}
\newcommand{\ben}{\begin{enumerate}}
\newcommand{\een}{\end{enumerate}}
\newcommand{\complex}{general }
\newcommand{\complexity}{generality }
\newcommand{\prior}{initial distribution }
\newcommand{\lrelax}{\lambda_{\rm relaxation}}
\newcommand{\tA}{$\tilde{A}$\,}
\newcommand{\tB}{$\tilde{B}$\,}
\newcommand{\tC}{$\tilde{C}$\,}
\newcommand{\mA}{${A}$\,}
\newcommand{\mB}{${B}$\,}
\newcommand{\mC}{${C}$\,}
\newcommand{\MS}{main~text\,}

\usepackage{hyperref}
\usepackage{graphics,graphicx,color}
\usepackage{bm}
\usepackage{amsmath}
\usepackage{verbatim}
\usepackage{multirow}
\usepackage{amssymb}
\usepackage{trfsigns}

\begin{document}
\title{Time-dependent information transmission in a model regulatory circuit}
\author{F. Mancini}
\email[]{fmancini@sissa.it}
\affiliation{International School for Advanced Studies (SISSA), Trieste, Italy}
\author{C. H. Wiggins}
\email[]{chris.wiggins@columbia.edu}
\affiliation{Department of Applied Physics and Applied Mathematics, Center for Computational Biology and Bioinformatics, Columbia University, New York, NY 10027}

\author{M. Marsili}
\email[]{marsili@ictp.it}
\affiliation{The Abdus Salam International Centre for Theoretical Physics (ICTP), Trieste, Italy}

\author{A. M. Walczak}
\email[]{awalczak@lpt.ens.fr}
\affiliation{CNRS and Laboratoire de Physique Th\'{e}orique de l'\'{E}cole Normale Sup\'{e}rieure, Paris, France. }

\date{\today}
\linespread{1}

\begin{abstract}
Many biological regulatory systems respond with a physiological delay when processing signals. A simple model of regulation which respects these features shows how the ability of a delayed output to transmit information is limited: at short times by the timescale of the dynamic input, at long times by that of the dynamic output. We find that topologies of maximally informative networks correspond to commonly occurring biological circuits linked to stress response and that circuits functioning out of steady state may exploit absorbing states to transmit information optimally.
\end{abstract}

\maketitle

\section{Introduction}

To respond to environmental changes, regulatory biochemical networks need to 
transform the molecular signals they receive as input into concentrations of response molecules. 
These processes are inherently stochastic, as both input and output molecules are often present in small numbers. 
This observation has motivated a number of recent works which pose the search for 
design principles as an optimization problem over the network topology and reaction rates. 
Not all designs of biochemical networks are equally represented in cells \cite{Alon_NatRevGen},
which raises the question of whether the prevalence of particular network motifs arises because they are best suited for specific tasks the cell has to fulfill.  
 In order to attempt to answer this question, and to explore the functions and limitations of given network architectures, 
 one can consider a well-defined objective function, such as rapidity of response \cite{Mangan2006, Alon_NatRevGen}, minimization of noise \cite{Saunders2009, Emberly08} or information transmission between the input and output \cite{TWB_PRE09, TWB_PRE10, TWB_PRE12, J-Phys-Cond-Matt_23_153102_2011}, and compare the performance of particular circuits under a set of constraints (e.g. molecular cost, noise). Phrasing the problem as an optimization over the parameters { and probability distributions} of these networks allows one to find the optimal circuit that best fits this one specific function. The optimal architecture corresponds to how one would ``design'' or ``build'' a circuit if the objective was to satisfy a known function. 

In this paper, we shall focus on optimal information transmission between an input and an output as a possible objective function. The idea of information optimization { in biomolecular networks} was tested in early fruit fly development, by {{maximizing the transmitted information over the input distribution and }}predicting the probability distribution of the hunchback protein that is regulated by the bicoid morphogen, using experimentally measured input-output relations and noise profiles \cite{PNAS_105_34_12265-12270_2008}. More recently, the information transmission measured in an inflammatory cytokine signaling pathway of murine fibroblasts \cite{Science_334_354_doi10.1126_2011} was used to find a tree-like network topology, thus allowing us to identify the bottlenecks 
that limit signal integration. This combined theoretical and experimental approach singled out the conditions under which  feedback, time integration and collective cell response can increase information transmission.  

Theoretical studies based on the optimization of information transmission in regulatory circuits have demonstrated that, within a network that functions at steady state, the system can exploit the molecular details of the network to transmit information while paying a molecular cost \cite{TWB_PRE09, WMW_PNAS09}. Positive feedback was shown to increase the effective integration time and average out input  fluctuations, thus allowing for reliable information transmission 
at low input concentrations \cite{TWB_PRE12}. On the contrary, negative feedback reduces noise at high input concentrations by reducing the effective nonlinearity of the input-output relation. 
Molecular strategies, such as using feed-forward loops \cite{TWB_PRE10} and slow binding-unbinding dynamics  \cite{MWW_PRE09}, also increase information transmission, because they lead to a quasi-binary readout and multimodal output distributions.

In many situations biochemical signals change with time, which has {led} to an interest
in the information-optimal response to pulses in signaling cycles \cite{Levine2007} and to oscillatory driving \cite{MWW_PRL10}. 
Similarly to what was found for stationary 
signals, circuits that produce bimodal output states \cite{MWW_PRL10} transmit more information. Tostevin and ten Wolde looked at time-dependent Gaussian processes in a linearized regulatory circuit and found that those network properties that are important for transmitting information about instantaneous signals may not be those that are relevant for information transmission of entire trajectories \cite{tostevintenwoldeprl, tostevintenwoldepre10}.
Within the same framework de Ronde and colleagues focused on understanding the role of regulation and found that positive feedback increases the fidelity of low-frequency signals, whereas negative feedback proves better at high frequencies \cite{derondetostevintenwoldepre10}. 
In  feed-forward circuits, they showed that topologies alone are not sufficient to characterize network properties, but that interaction strength also plays an important role \cite{derondetostevintenwoldepre12}.

In this paper, we focus on a similar set of issues, but we take into account that regulatory response is often at a delay relative to input signaling, because of e.g. transcription and translation processes, cellular compartmentalization, etc  \cite{AlbertsIIIedition}. 
Examples include the chemotactic response of bacteria \cite{Segall} or amoeba \cite{fullerpnas10} to nutrients or conversely to antibiotics \cite{LeBJ06}. This delay is intrinsic and unavoidable in a biological circuit - mRNAs and proteins are not produced instantaneously, and interpreting the initial signal takes time. Optimal design, therefore, entails maximizing information transmission between the input at a given time and the output at a later time. Nemenman \cite{Ilya} has shown that, in simple regulatory circuits, there is an optimal time lag for which the mutual information between the input and the delayed output is maximized.

In this paper we ask a different question. {\it Given a fixed unavoidable delay} in the response, what is the optimal way to design a regulatory system to optimally transmit information between the input at an initial time and an output at a later time? In our approach, the delay is an intrinsic property of the system, which we cannot control, but rather treat as an additional constraint. Given this constraint, we ask about the optimal parameters and architecture of the system that maximally transmits information between the input and the output. {{A system that transmits information optimally matches the properties of the input and output distributions with the regulatory relation of the network. To find this optimal system \cite{Shannon,Cover1991}, we optimize over both the {initial} 
distribution and the elements of the network. } }Our goal in this paper is to focus on how the natural cellular delays constrain the architectures of regulatory circuits. We find that the design of circuits that optimally transmit information between the input and a delayed output corresponds to known circuit's architectures (push-pull networks) \cite{Guisbert,LahavAlon,Brasier}. Moreover, we find that this prediction is robust, because optimal architectures do not change with the length of the imposed delay.

\begin{figure}[!ht]
\hspace{- 0.1\linewidth}
\includegraphics[scale=0.6]{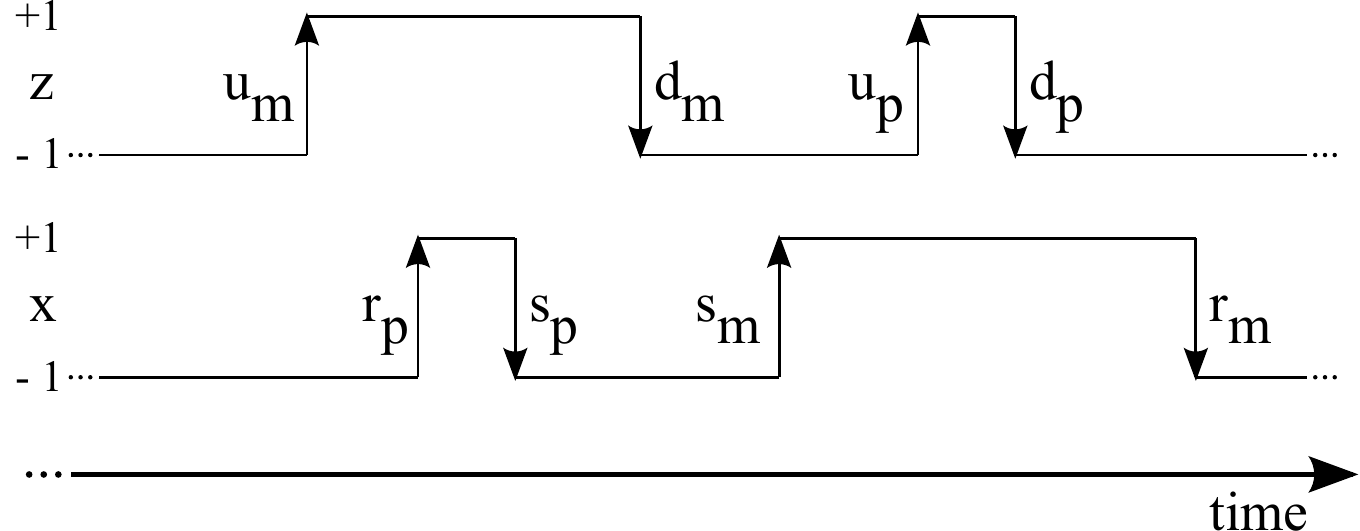}
\caption{\label{diagram} Time evolution of random variable $z(t)$, which models a biochemical input transitioning from/to a down-regulated state ($-1$) to/from an up-regulated state ($+1$), with rates $\{u_m,u_p\}$/$\{d_m,d_p\}$, respectively. Random variable $x(t)$ models activation ($+1$) or deactivation($-1$) of a biochemical output: it is regulated by $z$, with which it aligns (`activation', or up-regulation) with rates $r_m$ or $r_p$ or anti-aligns (`repression', or down-regulation) with rates $s_m$ or $s_p$. The subscripts $m$ and $p$ in the rates account for the state of the other variable, that is $-1$ and $+1$, respectively.}
\label{figdiag}
\end{figure}

Our strategy is to address these issues within the simplest possible model, composed of two binary components, $z$ and $x$, that switch randomly in continuous time (see Fig.~\ref{figdiag}). This can model, e.g., the expression state of two genes, or a gene and a protein, each of which can be up- or down-regulated. The model sacrifices molecular details of biochemical regulatory networks, which can be very complex and 
whose features can have an effect on their information-processing functions. Yet, the simplicity of our model makes our approach and the resulting results as clear as possible.

The paper is organized as follows: we start by introducing the simplified two-state model that allows us to optimize information given a fixed delayed response.   
We first consider the case in which the system is at steady state and the signal up-regulates (or down-regulates) the response with and without feedback. We then relax the steady-state assumption and jointly optimize the rates of the model and the \prior over the states of the system. We finish with a discussion of our results and limitations of the approach.

\section{Model description}

We consider a system of two dynamical variables $x, z$ that describe two genes, or a gene and a protein. Each one switches between an activated $+1$ and inactivated $-1$ state according to the rates defined in Fig.~\ref{figdiag}. {Specifically, $x$ aligns or anti-aligns with $z$ with rates $r$ or $s$, respectively. The subscripts $m$ and $p$ in the rates indicate the state of the other variable ($-1$ or $+1$) at the moment of the flip of the first variable.} 
Our system at any time is fully described by a four-state probability distribution $p(y)$, where $y{=}(x,z) \in \{(-, -), (-, +), (+, -), (+, +)\}$ is a joint variable for the output and the input. The temporal evolution of the conditional probability $p(y'|y)$ to find the system in state $y'$ at time $t$ given state $y$ at $t{=}0$ 
is given by a 
continuous-time master equation $\partial_t p{=} - \mathcal{L} p$, where $\cal{L}$ is a $4\times 4$ transition matrix set by the rates of switching between the states (defined in Fig.~\ref{figdiag} and constrained to be in the range $[0,1]$). In $p(y'|y)$ and in the rest of the paper, primed variables refer to the system state at time $t$, and unprimed variables to the initial time $0$. The solution of the master equation can be formally written as $p{=}e^{-t\mathcal{L}}$ and is conveniently expressed in terms of its expansion in left and right eigenvectors of $\mathcal{L}$ (see Appendix \ref{appA}). In particular, the (normalized) right eigenvector corresponding to the null eigenvalue is the stationary state $p_\infty(y'){=}\lim_{t\to\infty} p(y'|y)$.
 
As explained in the Introduction, for the sake of realism we want to take into account the intrinsic delays with which biochemical regulatory networks respond to signals. Therefore we compute the mutual information between the input $z$ at time $0$ and the output $x'$ at a time delay $t$, which is defined as
\begin{equation}
\label{Idef}
I[x_t,z_0]= \sum_{x',z} p(x',z)\log_2 \frac{p(x',z)}{p(x')p_0(z)}.
\end{equation}
The joint probability distribution $p(x',z)$ can be readily derived from the conditional distribution $p(y'|y)$ and the \prior $p_0 ( y )$ (See Appendix \ref{appA}). 

{{Intuitively, a system that conveys most information between the input and output requires matching the properties of the network (in the case of our model defined by the switching rates) with the properties of the input and output distributions. The maximum information transmitted by a system, termed the capacity of this system \cite{Shannon,Cover1991}, is defined as the optimum of Eq.~\ref{Idef}} with respect to the input distribution. We discuss two cases: in the first one the system, including the input to which it is responding, is in steady state, e.g. the response to a morphogen gradient such as fibroblast growth factor (FGF) in development \cite{Pourquie2003,Lawrence1992}. In the second case the initial state of the system is not the steady state, which triggers its response, e.g. production of an enzyme to metabolize a newly available sugar \cite{AlbertsIIIedition}.  }  While in the former the initial state of the regulatory network is determined by ${\cal{L}}$
(i.e. $p_0(y)=p_\infty(y)$), in the latter $p_0(y)$ may also be optimized. We consider both optimization cases in the context of our model.

{{A trivial way to maximize information transmission 
is for the input to change infinitely slowly relative to a fixed delay time $t$ (i.e. $u_l, d_l\to 0$, with $l=p,m$
following the notation of Fig.~\ref{figdiag}), such that for any finite $t$, the output yields a noiseless readout of the input, i.e. $p(x_t=z_0)\approx 1$. 
In short, nothing happens. Instead, we are interested in regulatory responses to changes in the input.}}

To constrain our optimization
such that information is transduced 
on a timescale set by the system's own dynamics, we optimize the quantity
\begin{equation}
\label{eqI}
\mathcal{I}(\tau)=I[x_{t=\tau/\lambda},z_0],
\end{equation}
where the rate $\lambda$ is given by the smallest non-zero
eigenvalue of ${\cal{L}}$ and is the inverse of the system's largest relaxation timescale. We remove arbitrariness in the choice of time units by fixing the magnitude of the maximum rate to be $1$. $\mathcal{I}(\tau)$ implicitly depends on the rates appearing in $\cal{L}$ 
and, if it is not set by $p_0(y)=p_\infty(y)$,  on the \prior $p_0(y)$. 
 We find network architectures that maximize ${\cal{I}}$ in Eq. (\ref{eqI}) for each  rescaled delay $\tau=\lambda t$ over the rates \footnote{Code that performs the optimization is available at http://infodyn.sourceforge.net.}. 

To summarize our procedure, for a fixed intrinsic delayed response of the output measured in units of the relaxation time of the system we look for the optimal rates of the network defined in Fig.~\ref{figdiag}. We then scan the delay time to see how the properties of the optimal networks change.

\section{Results}

\subsection{Simple activation}

To gain intuition, we start by considering the simplest possible case (model 
\mA) where $z$ up-regulates $x$
perfectly, symmetrically, and without feedback. In this case, 
$x$ is slaved to $z$ and switches only if $x\ne z$ but, due to the stochasticity of the model, it may not align immediately with $z$
(see Fig.~\ref{figdiag} with $s_m{=}s_p{=}0$, $u_m{=}u_p{=}d_p{=}d_m{\equiv}u$ and $r_p{=}r_m{\equiv}r$). This leaves us with only two timescales, related to $u$ and $r$. 

In the steady-state case ($p_0(y)=p_\infty(y)$), the mutual information can be computed explicitly and is related to the entropy of an effective two-state spin variable:
\begin{equation}
I[x_t,z_0] =\frac{1+\mu}{2}\log_2(1+\mu) +  \frac{1-\mu}{2}\log_2(1-\mu),
\label{infoent}
\end{equation}
where the ``effective magnetization'' 
$\mu{\equiv}2[p(x'{=}z,z)-p(x'{\neq}z,z)]$ is 
$( e^{-2 u t} r(r+2 u) - \e^{-r t}4 u r)/((r-2 u) (r+2 u))$ (see Appendix \ref{appD}) \footnote{For $r{=}2u$, $\mu{=}e^{-rt}/2(1-2rt)$.}.
For long times, we see that $\mu{\rightarrow}0$ and $I[x_t,z_0]{\rightarrow}0$, as expected. 

As previously shown for a different model \cite{Ilya}, the mutual information for a system initially in steady state has a maximum for a non-zero delay $t^*{=}{\log\left(\frac{2u+r}{2r} \right)}/{(2 u - r)}$, which is determined by the interplay of the two timescales introduced above. Here, we are not interested in finding the timescales over which information transmission is maximal, but rather the rates that maximize $\mathcal{I}(\tau)$ at {\em fixed}  \footnote{Note that, in un-rescaled time, maximal information transmission occurs for $u\to 0$ with $r$ finite, for any $t>0$, as $I[x_t,z_0]\to 1$.} $\tau=\lambda t$, where $\lambda=\min\{2u,r\}$ in model \mA ({optimal information curves are obtained as explained in} Fig.~\ref{hullfig}).

\begin{figure}[!ht]
\includegraphics[scale=0.65]{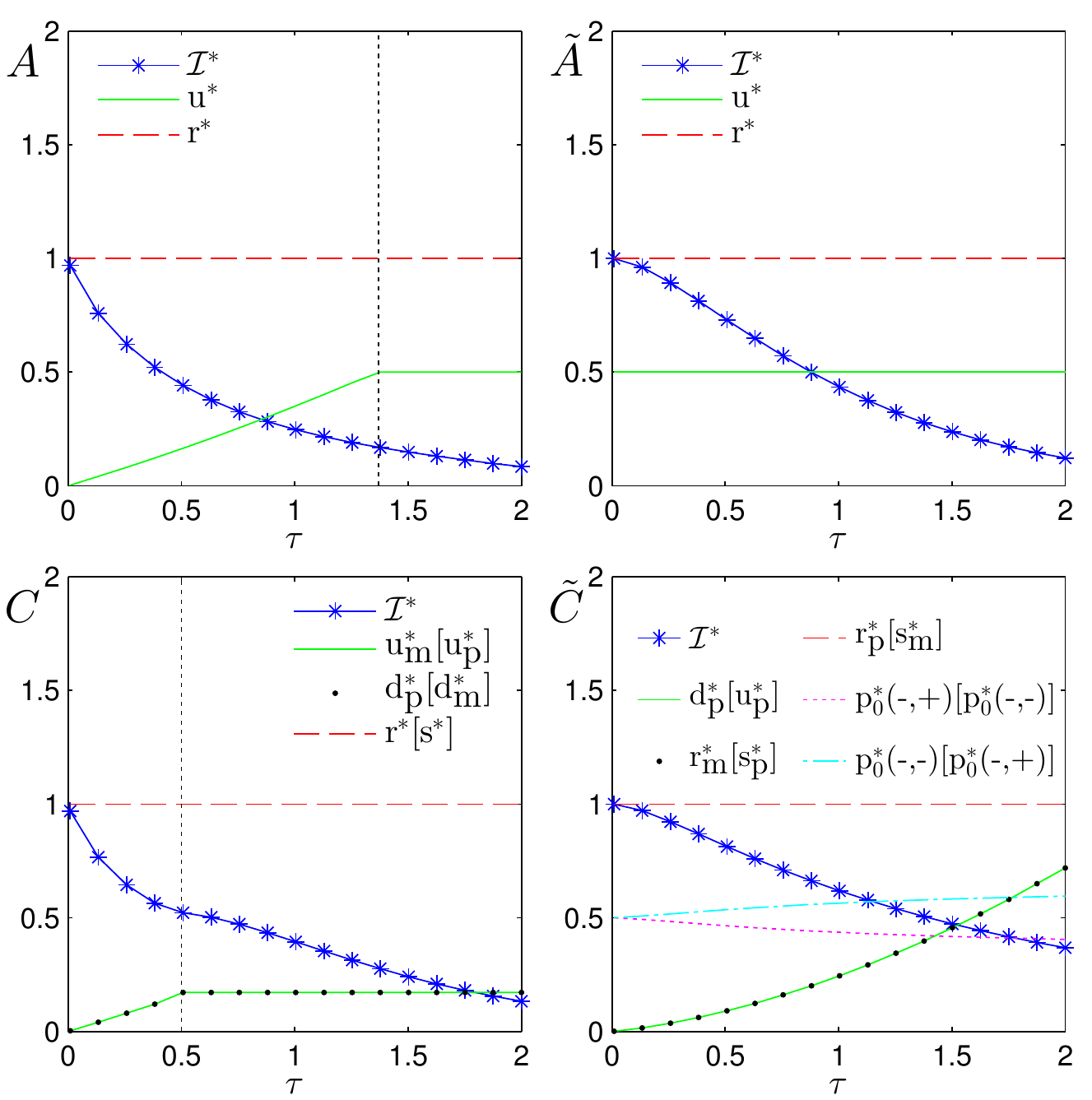}
\caption{\label{Fig2} Optimal parameters and transmitted information  $\mathcal{I}^*(\tau)$ for the activator and feedback models in steady state(\mA and \mC) and relaxing the steady-state assumption (\tA and \tC), as functions of the rescaled delay $\tau$. In panels \mC and \tC, parameters in square brackets refer to alternative optimal topologies (see right-hand diagrams in Fig.~\ref{FigPan} \mC and \tC). Subscripts $m,p$ are omitted when $x_m{=}x_p$ (with $x{=}u,d,r,s$). The results shown are valid for nonzero delays $\tau$.
Results for models \mB and \tB are shown in Appendix \ref{appG}.} 
\end{figure}

The optimal information $\mathcal{I}^*(\tau)$ and parameters $(u^*(\tau),r^*(\tau))$ for the simplest model where $z$ activates $x$ are plotted in Fig.~\ref{Fig2}~\mA. We see a clear crossover in terms of the switching rate $u^*$ that regulates the state of the input $z$ (see dashed vertical line in Fig.~\ref{Fig2} \mA): it initially increases in time and then plateaus at a value of $u^*{=}0.5$. This crossover results from the fact that for $r{>}2u$ the relaxation time is dominated by the rate at which the input changes ($\lambda{=}2u$), whereas for $r{\le}2u$ the rate at which the output changes fixes relaxation times ($\lambda{=}r$). Information transmission is dominated, over short timescales, by the faster rate $r$. Over long timescales, optimality is achieved by matching the characteristic times of the two processes, {that are equal to the inverse of the two smallest non-zero eigenvalues $\lambda_1{=}r$ and $\lambda_2{=}2u$}. The degeneracy of the two smallest non-zero eigenvalues for large $\tau$ is a non-trivial generic feature of optimal networks that we also find in more complex models (see below). The dynamics of model \mA can be summarized by the network topology shown in Fig.~\ref{FigPan} \mA.

\subsection{Activation/repression}

We can generalize model~\mA by allowing $z$ to regulate $x$ asymmetrically -- that is, $r_p\neq r_m$ -- and to down-regulate it as well 
--- that is, to allow $s_m, s_p{\neq}0$ (model \mB).
As in model \mA, we forbid
feedback from $x$ to $z$, hence the transition rates for $z$ do not depend on the state of $x$ (i.e. $u_m{=}u_p{\equiv}u$ and $d_p{=}d_m{\equiv}d$). Optimization yields
only solutions coinciding with that of model \mA, or with its symmetric counterpart (wherein $z$ perfectly down-regulates $x$ instead of perfectly 
up-regulating: $r_p{=}r_m{=}0$ and $s_p{=}s_m{\equiv}s$). {Results are shown in Fig.\ref{modelB} and \ref{FigPan_SI} of Appendix \ref{appG}.}
Intuitively, in order for information to be transduced between $x$ and $z$,
they either align or anti-align, resulting in the common simple activator or repressor element \cite{Alon_NatRevGen}. 
Note that the same topological structure is found at all timescales $\tau$. This is to be contrasted with previous studies \cite{WMW_PNAS09, TWB_PRE09} which, taking into account the molecular cost paid by producing higher copy number (e.g., creating more proteins), have found small discrepancies between the information transmitted in the two cases of up-regulation and down-regulation. Since our model does not explicitly account for protein copies, we do not observe such a difference. 

\subsection{Role of feedback}
\label{feedback}

\begin{figure}[!ht]
\includegraphics[scale=0.65]{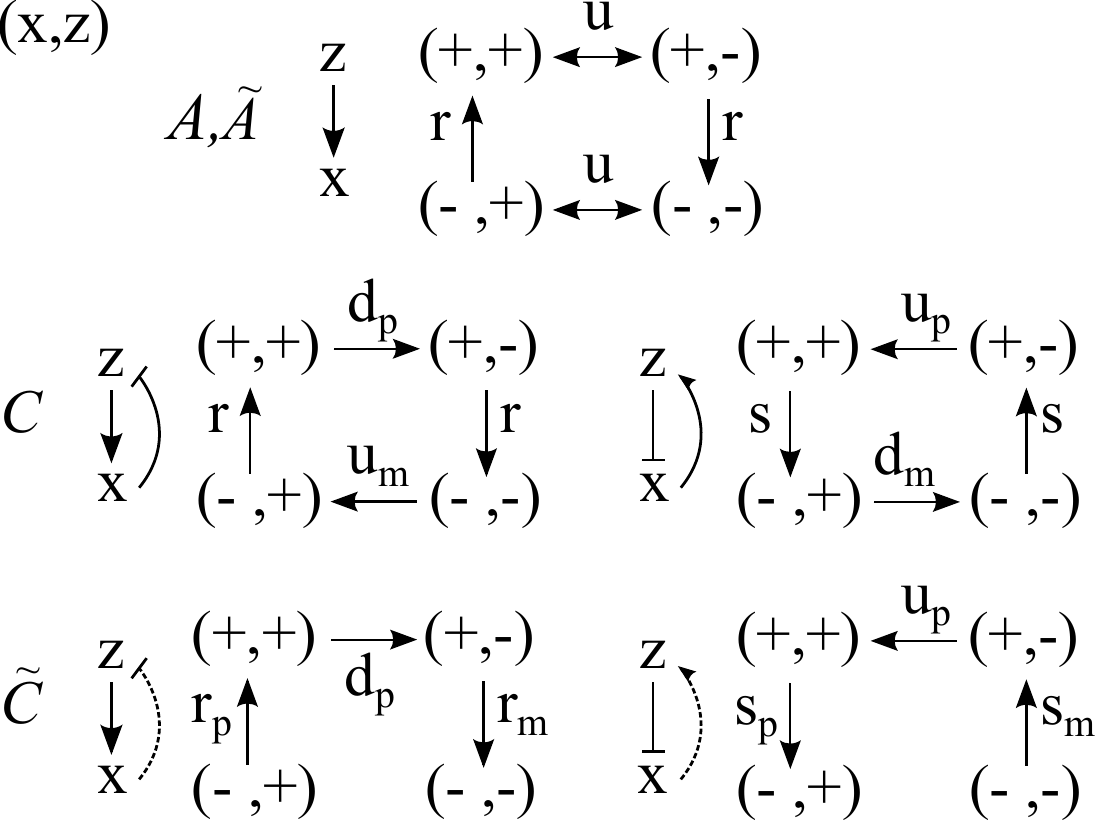}
\caption{ Optimal topologies and dynamics for the activator and feedback models in steady state (\mA and \mC) and relaxing the steady-state assumption (\tA and \tC). In the two feedback models (\mC and \tC) the optimal topology is a ``push-pull'': one gene (or protein) activates the other, which in turn represses the first gene/protein (as we can see, the roles of $z$ and $x$ are interchangeable). The dashed line in \tC means that the feedback exists only until the absorbing state is reached. The topologies of these optimal networks were found by inspecting the optimal rates manually. Results for models \mB and \tB are shown in Appendix \ref{appG}.}
\label{FigPan}
\end{figure}

Recent studies \cite{TWB_PRE12, derondetostevintenwoldepre10,derondetostevintenwoldepre12} have pointed to the important role of feedback in transmitting information, a form of which we can consider using the full set of 8 rates in Fig.~\ref{figdiag} (model C).
Now the hierarchical relation between $z$ and $x$ is broken: both can regulate each other's expression, either by down- or up-regulation. Examples of maximally informative 
topologies for all possible rescaled delays $\tau$ 
are illustrated in Fig.~\ref{FigPan} \mC and 
reveal a ``push-pull'' network - one gene (or protein) up-regulates the other, which in turn down-regulates the first gene or protein. Such push-pull circuits are very common in biology,
from microbes \cite{Guisbert} to humans (\cite{LahavAlon} and references therein) as a source of oscillations \cite{Hasty} and pulse responses \cite{LahavAlon} (see discussion below). 
Again, due to the symmetry of the problem, 
we can flip either $x$ or $z$ and we find 4 equally informative solutions (two of which are shown in Fig.~\ref{FigPan} \mC), associated with different sets  of the 8 rates being driven to zero by the optimization procedure. { We exploit the numerical observation that certain rates are zero {to} simplify the matrix $\cal{L}$ and find an analytical expression for the information and the optimal rates {(see Appendix \ref{appE})}.}  We find that the optimal value of the input rates ($u_m[u_p], d_p[d_m]$) now plateaus at a value of $3-2\sqrt{2}$ for $\tau>\tau^*=0.5$ (see dashed vertical line in Fig.~\ref{Fig2} \mC): this value is set by competition with the $r[s]$ rate, by matching the two smallest non-zero eigenvalues in order to avoid oscillatory solutions (for input rates $>3-2\sqrt{2}$, the eigenvalues become imaginary describing oscillations; see Appendix \ref{appE} for a more detailed derivation). Push-pull networks can oscillate \cite{Hasty} \footnote{Delay in mRNA production was shown to be a necessary element of stable oscillations, making them hard to observe in synthetic networks \cite{Hasty}.}, thwarting optimal information transmission by decorrelating the system, hence the oscillatory regime is not the optimal solution.

The optimal solution exploits feedback to transmit more information. For small delay times, feedback does not play a role and models \mA and \mC transmit the same amount of information. For delay times at which the input rates have reached their plateau value, the optimal circuit of model \mC can transmit more information at a fixed $\tau$ than the optimal circuit without feedback of model \mA (Fig.~\ref{Fig2}). Intuitively, this happens because the value of the switching rates of $z$ at the plateau is smaller in model \mC than in model \mA, thus $z$ is less likely to switch on/off. As a result of these slower switching rates, the system in model \mA is more likely to cycle through the four states and hence to obscure correlations with the initial condition than in model \mC. 
Additionally, feedback leads to a rotational directionality among the transitions that is not observed in model A
(cf. Fig.~\ref{FigPan}). As a result of this rotational directionality the system never directly `flips back',
enhancing the transduced information. In summary, feedback allows for a combination of slower flipping rates and imposed order to the visited states that enables {to read out more information about the initial state at later times.}

\subsection{{ Systems out of steady state}}
\label{optimalp0}

Having discussed optimal delayed information transmission of repeated readouts in the stationary state, we
now consider regulatory  networks that optimize a one-time response to an input (e.g. by producing an enzyme to metabolize a nutrient that appeared in the environment). {{We assume that at time $t_0$, the system is in an initial state which is not its steady state. We ask what is the optimal design of the circuit to produce the most informative output  to this initial state given a fixed delay. }}
Unlike in the previous case where the network was in steady state (which determined the initial distribution) and would respond repeatedly, we now also ask what is the optimal initial distribution of the system. We allow the initial and final distribution to be different. 
{{The optimization over the input distribution describes the matching of the properties of the regulatory network and the initial non-stationary state corresponding to the e.g. appearance of a sugar source in the cell, DNA damage or food shortages.  }}
Specifically,
we consider the same three models studied above (\mA, \mB, \mC), 
but now we optimize simultaneously not only on the rates but also on the initial probability distribution $p_0(y)$ and refer to the associated models as \tA, \tB  and \tC, respectively.

{ To calculate the capacity of the system, we optimize over the initial distribution and the parameters of the network. We could fix the initial probability distribution and optimize only over the network parameters. However this would be an arbitrary choice of the initial distribution and we would not calculate the capacity of the system. Instead, we consider all possible input distributions and ask which one of them guarantees maximal information transmission. Cells often pre-process external signals, for example the lack of glucose is presented in terms of high cAMP (or more specifically activated crp) concentration to the lac operon, which gives the cell a certain degree of control over the distribution of the input to a network}.

We start with model \tA, which enjoys an up-down symmetry and suggests parameterizing the \prior $p_0(y){=}p_0(x,z)$ via $p_0(+,+){=}p_0(-,-){=}(1{+}\mu_0)/4$ and $p_0(+,-){=}p_0(-,+){=}(1{-}\mu_0)/4$~\footnote{The case of a steady-state \prior is a special case of a system that enjoys this property.}. The mutual information is still given by Eq.~\ref{infoent}, with $\mu{=}\mu_0 e^{-rt} + \frac{r}{r-2u} (e^{-2u t}-e^{-rt})$. First we consider the properties of mutual information as a function of time, without fixing the delayed readout. For $t{=}0$, the highest information is attained for $\mu{=}\mu_0{=}\pm 1$, when $z$ and $x$ are perfectly aligned/anti-aligned. Moreover, $\mu$ and $I[x_t, z_0]$ decrease exponentially with time $t$. Therefore, unlike in model \mA, the information transmission does not improve by making a delayed readout. In other terms, the absence of a maximum at $t^*>0$ in $I[x_t,z_0]$ for optimal initial states suggests that, at odds with the stationary case (model \mA), the mechanism for information transmission is only 
governed by the loss of information about the initial state as the system relaxes to stationarity. 

After performing the optimization of $\mathcal{I}(\tau)$, we find that for each rescaled delay $\tau$ the optimal \prior $p_0(y)$ concentrates on the states $(+,+)$ and $(-,-)$. We can understand this result intuitively: the rate for switching out of these states, $u$, is small, so the system is more likely to remain in these states than in the other two states (see Fig.~\ref{Fig2}\tA and Fig.~\ref{FigPan}\tA). Posing the system in these long-lived states allows for more information transmission about the initial condition at small readout delays $\tau$.

We now turn to maximizing information transmission over the \prior $p_0(y)$ in the more \complex models \tB and \tC. As above, symmetry provides a number of optimal networks related by permutations (see
Fig.~\ref{FigPan_SI}\tB and Fig.~\ref{FigPan_SI_Ctilde}). The optimal rates are shown for the case of $z$ up-regulating [down-regulating] $x$ in Fig.~\ref{modelB}\tB and Fig.~\ref{Fig2}\tC.
We find a qualitative difference in design as compared to the stationary case (models \mB and \mC): while the optimal topology remains the same, now either one of the aligned or non-aligned states becomes absorbing, e.g., $p_\infty(y'){=}\delta_{y',(+,+)}$ or $p_\infty(y'){=}\delta_{y',(-,+)}$ in the examples in Fig.~\ref{FigPan}\tC \footnote{In each of the $4$ degenerate topologies, a different state becomes absorbing.}. 

The occurrence of an absorbing state, with a nearly-equal
optimal \prior $p_0(y)$ over the initial and final states, limits the system's dynamics and
leads to the optimal topology for a one-time response. 
In the absence of feedback (model \tB, e.g.~receptor activation in a complex pathway), when the system, initially in the inactive state $(x,z)=(-,-)$ is presented with a signal $(x,z)\to (-,+)$, it switches on a response $(x,z)\to (+,+)$ (see Fig.~\ref{FigPan_SI}\tB). However, in the presence of feedback ({model \tC}, e.g. a nutrient activating the production of an enzyme for its uptake, amino acid biosynthesis) the optimal dynamics includes ``feedback inhibition", in which the output switches off the input (see Fig.~\ref{FigPan}\tC) \cite{AlbertsIIIedition}\footnote{Such design is reminiscent of Shannon's Ultimate Machine: a device with a switch that, when it is turned on, activates an arm that switches it off.}. As in model \mC, feedback imposes an order to the visited states, with a smaller rate for $z$ transitions than for $x$ transitions: these two features allow again for higher information transmission about the initial state (see Fig.~\ref{Fig2}\tC).

\indent
Let us consider a specific biological example - lactose metabolism - for the optimal network shown in Fig.~\ref{FigPan}\tC (left panel) and its corresponding optimal rates and \prior presented in Fig.~\ref{Fig2}\tC. The system describes two elements: lactose ($z$) and the degrading enzyme beta-galactosidase in the lac operon, ($x$). Our optimal solution consists of the rates and topologies of the network and the initial probability distribution of the system. The optimal solution is based on the matching between the statistics of the input and the output (lactose and beta-galactosidase) and the properties of the network, similarly to the approach taken in neuroscience by Laughlin \cite{Laughlin} and in studying time independent models of information transmission in molecular systems \cite{TWB_PRE09, TWB_PRE10, TWB_PRE12, WMW_PNAS09}. We find that the optimal initial distribution limits the system to be with close to equal probability in either the $(x,z)=(-,+)$ or the $(x,z)=(-,-)$ state, which corresponds to states in which the enzyme is turned off and the lactose is either there or not. In the example of sugar metabolism, these initial states are implemented using the lac repressor, that senses whether lactose is present or not in the environment and represses the lac operon when needed.

Our calculation does not fix the initial state of the network, but assigns initial probabilities to all states. This network design is optimal given this nearly equal probability of the input sensing sugar in the environment {(for example, activated crp in response to glucose in the lac operon)}. However, we did not constrain the {initial} distribution, but we asked for a best matching between the properties of the circuit and any initial distribution. If the probability distribution of the sugar in the environment were known and fixed, we would have to optimize the network given this additional constraints. The calculation presented in models \mA, \mB and \mC is a specific example of when the input probability distribution is constrained to be the steady-state distribution of the system (and therefore uniquely set by the optimal rates).

According to our optimal solution, lactose can either be present or not in the cell. In both cases the degradation enzymes are switched off. If there is no lactose, the system at $t_0$ is in the final absorbing state $(x,z)=(-,-)$. If at $t_0$ lactose ($z$) is sensed, the enzymes $x$ that degrade sugar are switched off and the system is in the $(x,z)=(-,+)$ state. The appearance of lactose activates the synthesis of  enzymes ($(-,+)\rightarrow (+,+)$), which cause the depletion of the sugar  ($(+,+)\rightarrow (+,-)$). Finally the lack of lactose de-activates the enzymes and the enzymes are degraded ($(+,-)\rightarrow (-,-)$). This matching between the initial probability distribution of seeing sugar in the environment and the regulatory elements of the network allows the system to transmit most information about the original state of the input with a delay. Given the fast initial rate for leaving the  $(x,z)=(-,+)$, if the enzyme is present, then sugar was initially present. If there is no enzyme, there was no sugar \footnote{In a real system the enzymes could be present at a basal level even with the absence of sugar, but our coarse-grained model cannot account for that.}.

The signal can directly be the input, as in the lactose metabolism example, or it can influence the input.  A biological example of the push-pull network shown in Fig.~\ref{FigPan}\tC (left panel) with the external signal that triggers an input is the p53-MdM2 circuit that is involved in DNA repair \cite{LahavAlon, Mayo2002}. The tumor suppressor protein p53 transcriptionally activates the MdM2 gene, the product of which degrades p53. DNA damage leads to an increase in p53 levels ($(x,z)=(-,+)$), which in turn up-regulates MdM2 ($(-,+)\rightarrow(+,+)$) that degrades p53  ($(+,+)\rightarrow(+,-)$) and in turn down-regulates  MdM2 ($(+,-)\rightarrow(-,-)$). In our optimal solution, the initial distribution of this network is tuned to a roughly equal probability of there being DNA damage or not. \\
In both the lactose and DNA damage case, the optimal networks perform one readout, after which they need to be reset externally: additional sugar needs to be taken up from the environment or p53 levels need to be increased by new DNA damage.

\begin{figure}[!ht]
\includegraphics[scale=0.65]{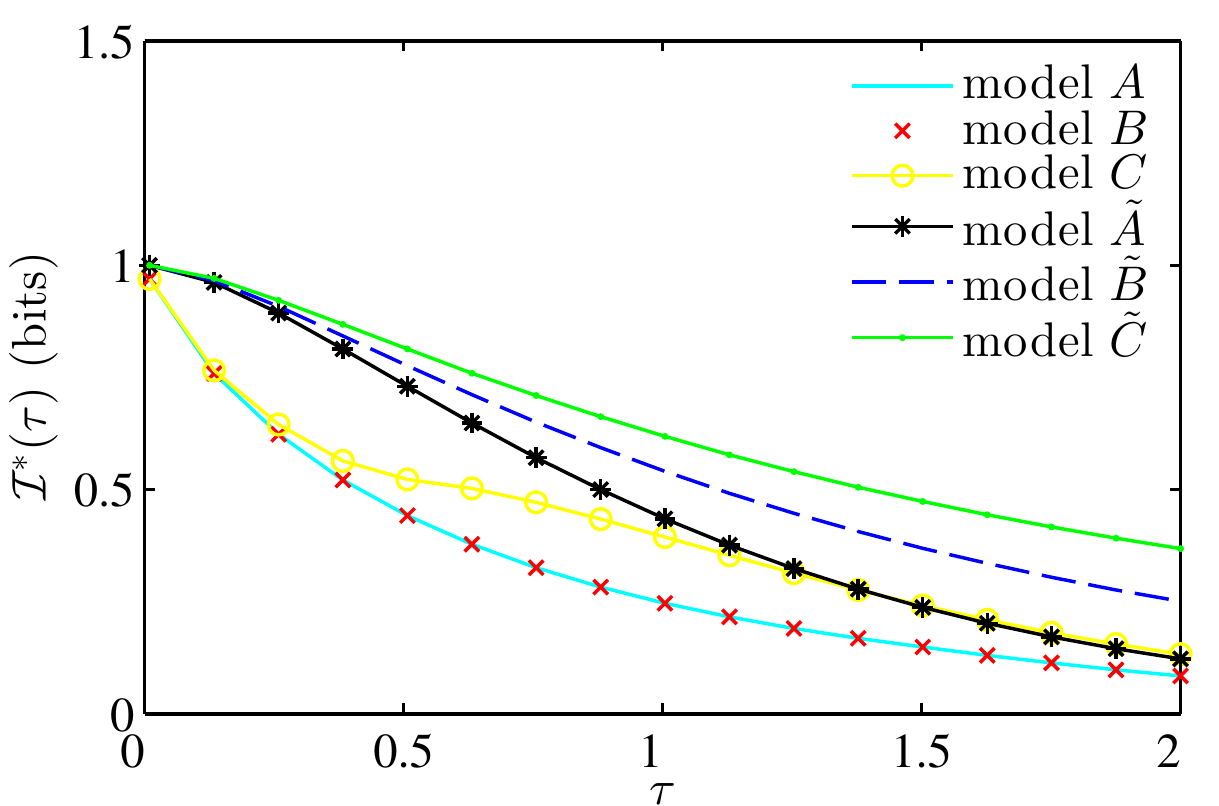}
\caption{\label{fig:allcases} Summary of optimization results: for each of the six models, the hull curve of the maximized information $\mathcal{I}(\tau)$ is plotted versus $\tau$. When feedback is present (model \mC) information is higher for $\tau \gtrsim 0.4$. When the system is initially in an optimal state (model \tA, \tB, \tC) the information is higher for each $\tau$ and its time decay is qualitatively different.}
\label{ALLINFO}
\end{figure}

\section{Discussion}

In Fig.~\ref{ALLINFO} we plot a comparison of all the cases possible within the two-state model. As the model \complexity increases (from \mA to \tC), so does the number of parameters; accordingly, the information capacity of the system also increases. As explained in Section \ref{feedback}, we see that the introduction of feedback in model \mC does not play a role in increasing information transmission for small $\tau$. However, the information gain coming from feedback is substantial for long rescaled delays $\tau$ between the input and output readout. 
Information transmission can be improved beyond that achieved in the steady state (\mA, \mB and \mC) if the system is pushed out of equilibrium in specific ways (\tA, \tB and \tC) to respond to one time signals (for details see Section \ref{optimalp0}). This increase in information is achieved by simultaneously optimizing the initial distribution of inputs {and outputs} in a way that matches the properties of the regulatory circuit.

Both in the steady-state and the non-steady-state feedback models, we find the optimal network topology of a push-pull circuit. Such circuits exists in many cells, ranging from bacterial (heat-shock response \cite{Guisbert}) to mammalian (I-$\kappa$B-NF$\kappa$B circuit \cite{Brasier} in animal stress response, p53-Mdm2 network involved in DNA damage response \cite{LahavAlon}), and they often combine a slow (transcriptional regulation) with a fast (protein-protein interaction) component, similarly to the design of our optimal architectures. In particular, the non-steady-state optimal topologies feature absorbing states, which result in single pulse responses. These responses are common in the case of stress signals and in some cases feature a ``digital'' behavior: the number of pulses, rather than the response intensity, is proportional to the input strength \cite{LahavAlon}. 

We find that allowing for feedback (models \mC and \tC) results in optimal solutions that transmit more information than models without feedback. This observation is similar to those found for instantaneous information transmission \cite{TWB_PRE10} and for the rate of information transmission \cite{derondetostevintenwoldepre10}. Our optimal solutions consist of two elements: the combination of slow and fast {rates} of reactions and the imposed rotational order of the states. In the non-equilibrium circuits where detailed balanced can be broken, the order of visited states is further enforced by the absorbing state, which means each state can only be visited once. These two general design elements make the readout of the input from the output more distinguishable, which in turn increases the amount of transmitted information, similarly to what was previously noted for more detailed molecular models and instantaneous information transmission. As noted, many push-pull networks that take part in stress response have the slow and fast  timescale that are encoded in a circuit that combines protein-protein interactions (fast) with gene regulation (slow timescales). Other molecular implementations of these general principles may be possible, however this simple model points to very general design elements. 

By studying simple two-state models of biochemical systems, we cannot interpret our optimal circuits in terms of the specific molecular designs that could be used to implement these networks. Specifically we do not account for molecular noise that comes from discrete numbers of proteins, mRNAs and genes in a regulatory circuit and that has been shown to play an important role in choosing certain regulatory elements over others \cite{WMW_PNAS09, TWB_PRE09}. The role of molecular noise and cost in the design of circuits that transmit information at a delay needs to be examined using more detailed models. The results presented here can be used as a starting point. 

In our calculation we do not explicitly model fluctuations in the signal, such as was done in previous work that considered information of an instantaneous response \cite{Levine2007} or looked at optimal delay times \cite{Ilya}. We just look at the optimal network that would best respond with a delay to a change in a signal. Due to this formulation we do not study fluctuations in the signal and we cannot address the question of whether the network is able to distinguish random (``irrelevant'') fluctuations from a real change in the signal - a question that is very important in understanding the design of biological circuits. 

This study gives a framework for studying information transmission in biochemical regulatory systems subjected to delays, also in non-stationary conditions. Such an approach can be extended to more realistic models that explicitly account for protein concentrations, where costs of protein production and degradation can be studied in detail. 

\section*{Acknowledgements}
We thank William Bialek for helpful discussions. AMW is funded by a Marie Curie Career Integration Grant. 

\appendix

\section{Calculating mutual information}
\label{appA}
In the \MS we calculate the mutual information between the input $z$ at time $0$ and the output $x'$ at a time delay $t$ using the temporal evolution of the joint probability distribution $p(x',z)$ obtained from a master equation. In this Appendix we give a detailed derivation of the steps of this calculation.

We define the state $y$ of the system as
\begin{equation}
y = (x,z) \in \{(-, -), (-, +), (+, -), (+, +)\},
\end{equation}
and the dynamics in terms of transition rate matrix
\begin{equation}
\mathcal{L} = \left( \begin{array}{cccc}
                      u_m+s_m & -d_m & -r_m & 0 \\
		      -u_m & d_m+r_p & 0 & -s_p \\
		      -s_m & 0 & u_p+r_m & -d_p \\
		      0 & -r_p & -u_p & d_p+s_p
                     \end{array}
	      \right).
\end{equation}
The corresponding master equation is 
\begin{equation}
\frac{dp}{dt} = - \mathcal{L} p
\end{equation}
for a vector $p{=}p(y){=}(p(-,-),p(-,+),p(+,-),p(+,+))$ (we omit the
    implied dependence on time). 
    Primed variables (e.g., $y'$) refer to the state of the
    system at time $t{\ne}0$; unprimed ones refer to the state at $t{=}0$.
    
The transition probability matrix $p(y'|y)$ is a solution of the master equation with initial condition 
\begin{equation}
\lim_{t\to 0}p(y'|y)=\delta_{y',y}
\end{equation}
and it can be written as the $(y',y)$ element of the matrix $\e^{-t\mathcal{L}}$, i.e.
\begin{equation}
\label{eqn:pxzspectral}
 p(y'|y) = \left[ \e^{-t\mathcal{L}}\right]_{y',y} = \sum_{\alpha=1}^{4} 
 \e^{-\lambda_{\alpha} t} v_{\alpha}(y')
 u_{\alpha}(y).
\end{equation}
$\lambda_{\alpha}$ (with $\alpha=1,\dots,4$) are the four (assumed to be distinct for this derivation) eigenvalues of $\mathcal{L}$, and $v_{\alpha}$ and $u_{\alpha}^{T}$ are their corresponding orthonormal right and left eigenvectors, with components $v_\alpha(y')$ and $u_\alpha(y)$:

\begin{eqnarray}
 \mathcal{L} v_{\alpha} &=& \lambda_{\alpha} v_{\alpha},\\
 u_{\alpha}^{T} \mathcal{L}  &=& u_{\alpha}^{T} \lambda_{\alpha}, \\
 u_\alpha^Tv_\beta&=&\delta_{\alpha,\beta}.
\end{eqnarray}
In particular, if we choose a normalization such that $u_1=(1,1,1,1)$, the eigenvector $v_1$, corresponding to the eigenvalue $\lambda_1=0$, is the stationary state 
$p_{\infty}(y)$. 
We are computing the mutual information between $z$ at time $0$ and $x'$ at time $t$, which is a function of $t$ and is given by
\begin{equation}
I[x_t,z_0]= \sum_{x',z} p(x',z)\log_2 \frac{p(x',z)}{p(x')p(z)}.
\end{equation} 
The joint probability $p(x',z)$ is calculated from $p(y'|y)$ as
\begin{widetext}
\beq
p(x',z)=&\sum_{y,y'}p(x',z|y',y)p(y',y)&
  {\rm using~the~definition~of~conditional~probabilities}\notag\\
=&\sum_{y,y'}p(x'|y')p(z|y)p(y',y)&{\rm exploiting~the~conditional~independence~of~}x',z \notag\\
=&\sum_{y,y'}p(x'|y')p(z|y)p(y'|y)p_0(y)&
  {\rm using~the~definition~of~conditional~probabilities}.
\label{eqn:pxz}
\eeq
\end{widetext}
Note that the elements of $p(x'|y')$ and $p(z|y)$ are either $0$ or $1$
according to whether, for example,  $y$ is consistent or inconsistent with $z$:

\beq
p(z=+|y=(+,+))&=&1\notag\\
p(z=+|y=(+,-))&=&0\label{eqn:zeroone}\\
\vdots\notag
\eeq
et cetera.
Finally, the marginal probabilities $p(z)$ and $p(x')$ are given by
$$
p(z)\equiv \sum_{x'} p(x',z), \quad p(x')\equiv\sum_z p(x',z).
$$

The numerical computation of the mutual information can now be implemented and the optimal rates for systems of various complexity can be found numerically. In the paragraphs below we present useful computational details for implementing this calculation. We have implemented the optimization procedure in MATLAB and we made the source code available via
the following public access repository: http://infodyn.sourceforge.net.

For certain models we can also make analytical progress by exploiting spectral representations of the joint distribution, as shown in Appendices~\ref{appD},~\ref{appE} and ~\ref{appF}. 

{\bf Numerical computation of the joint distribution.} For numerical computation in MATLAB, it is useful to rewrite Eq. \ref{eqn:pxz}
in terms of matrix operations. To that end, we define
(note that $X$ and $Z$ are $0-1$ matrices -- whose elements are $0$ or $1$, as per Eq. \ref{eqn:zeroone})
\beq
X_{x',y'} &\equiv & p(x'|y')\\
G_{y',y_0}&\equiv & p(y'|y_0)=
\left[ \e^{-t\mathcal{L}}\right]_{y',y_0} 
\\
P_{y_0,y} &\equiv & p_0(y_0)\delta_{y_0,y}\\
Z_{y,z}   &\equiv & p(z|y).
\eeq
This allows us to write Eq. \ref{eqn:pxz} compactly as
\beq
p(x',z)&=&\sum_{y'} X_{x,y'}\sum_{y_0} G_{y',y_0}\sum_y P_{y_0,y} Z_{y,z}\\
&=&\left[ XGPZ\right]_{x',z}
\label{eqn:pxzmat}
\eeq
that is how the equation is implemented in MATLAB.

{\bf Analytical calculation of the joint distribution.} For analytic calculations, it is useful to expand $p(y'|y)$ in Eq. \ref{eqn:pxz}
in terms of its spectral representation (Eq. \ref{eqn:pxzspectral}):
\begin{widetext}
\beq
p(x',z)&=&\sum_{\alpha}\sum_{y',y}p(x'|y')p(z|y)\e^{-\lambda_\alpha t} v_\alpha(y') u_\alpha(y) p_0(y)\notag\\
&=&\sum_{\alpha}
  \e^{-\lambda_\alpha t} 
  \left(\sum_{y'}p(x'|y')v_\alpha(y') \right)\!\!\!
  \left(\sum_{y }p(z|y) u_\alpha(y) p_0(y)\right)\notag\\
&=&
  \!\!\! \left(\!\! \sum_{y'}p(x'|y')p_\infty(y') \right)\!\!\!
  \left(\!\! \sum_{y }p(z|y) p_0(y)\right)+
\sum_{\alpha>1}
  \e^{-\lambda_\alpha t} 
   \left(\!\! \sum_{y'}p(x'|y')v_\alpha(y') \right)\!\!\!
   \left(\!\! \sum_{y }p(z|y) u_\alpha(y) p_0(y)\right)\notag\\
&\equiv&
  p_\infty(x')p_0(z)+
\sum_{\alpha>0}
  \e^{-\lambda_\alpha t}
  {\tilde{v}}_\alpha^{x'}
  {\tilde{u}}_\alpha^{z}
\eeq
\end{widetext}
where
\beq
  p_\infty(x')&\equiv& \sum_{y'} p(x'|y')p_\infty(y')\\
  p_0(z)&\equiv& \sum_{y} p(z|y)p_0(y)\\
  {\tilde{v}}_\alpha^{x'}&\equiv&\sum_{y'}v({y'})_\alpha p(x'|y')\\
  {\tilde{u}}_\alpha^{z}&\equiv&\sum_{y} u_\alpha(y) p_0(y) p(z|y).
\eeq
Writing $p(x',z)$ in this form makes it clear that, if the eigenvalues are distinct
and thus $p(y'|y)$ is diagonalizable, then $p(x',z)$ factorizes 
as $t\rightarrow \infty$ and thus $I[x_t,z_0]\rightarrow 0$.
Also clear is that $p(x',z)-p_\infty(x')p_0(z)$ is expressible as a sum of time-decaying
exponentials. Since $p(x'|y')$ and $p(z|y)$ are $0-1$ matrices,  in many cases $\tilde{v}_\alpha$ and $\tilde{u}_\alpha$ can be calculated explicitly, as shown below.

\section{Optimization procedure}
\label{appB}

We optimize 
over the parameters of each model in order to maximize $\mathcal{I}(\tau)=I[x_{t=\tau/\lambda},z_0]$, where $\tau$ is a dimensionless quantity that results from the rescaling procedure:
$$t\rightarrow t \cdot \lambda \equiv \tau,$$
where $\lambda$ is the inverse of the system's largest relaxation rate (the smallest nonzero eigenvalue of the rate matrix $\mathcal{L}$).
The steps of the ``rescale and optimize'' procedure are:
\ben
\item[]while $\tau_{min}<\tau<\tau_{max}$:
\item ~~optimize $\mathcal{I}(\tau;\theta)$ over parameters $\theta$ or parameters $\theta$ and \prior $p_0(y)$ 
\item ~~save $\mathcal{I}^*, \theta^*, p_0^*$
\item ~~increment $\tau$
\item[]end loop over $\tau$
\een
where
\ben
\item[]calculate $\mathcal{I}(\tau;\theta)$:
\item     calculate $\mathcal{L}(\theta)$
\item     calculate $\lambda(\mathcal{L})$
\item     calculate $p(x,z){=}X \exp(-\tau \mathcal{L}/\lambda) P Z$, as in Eq.~\ref{eqn:pxzmat}
\item     calculate $I[p(x,z)]$
\item[] return $I[p(x,z)]$ to optimization algorithm
\een

\begin{figure}[!ht]
\includegraphics[scale=0.6]{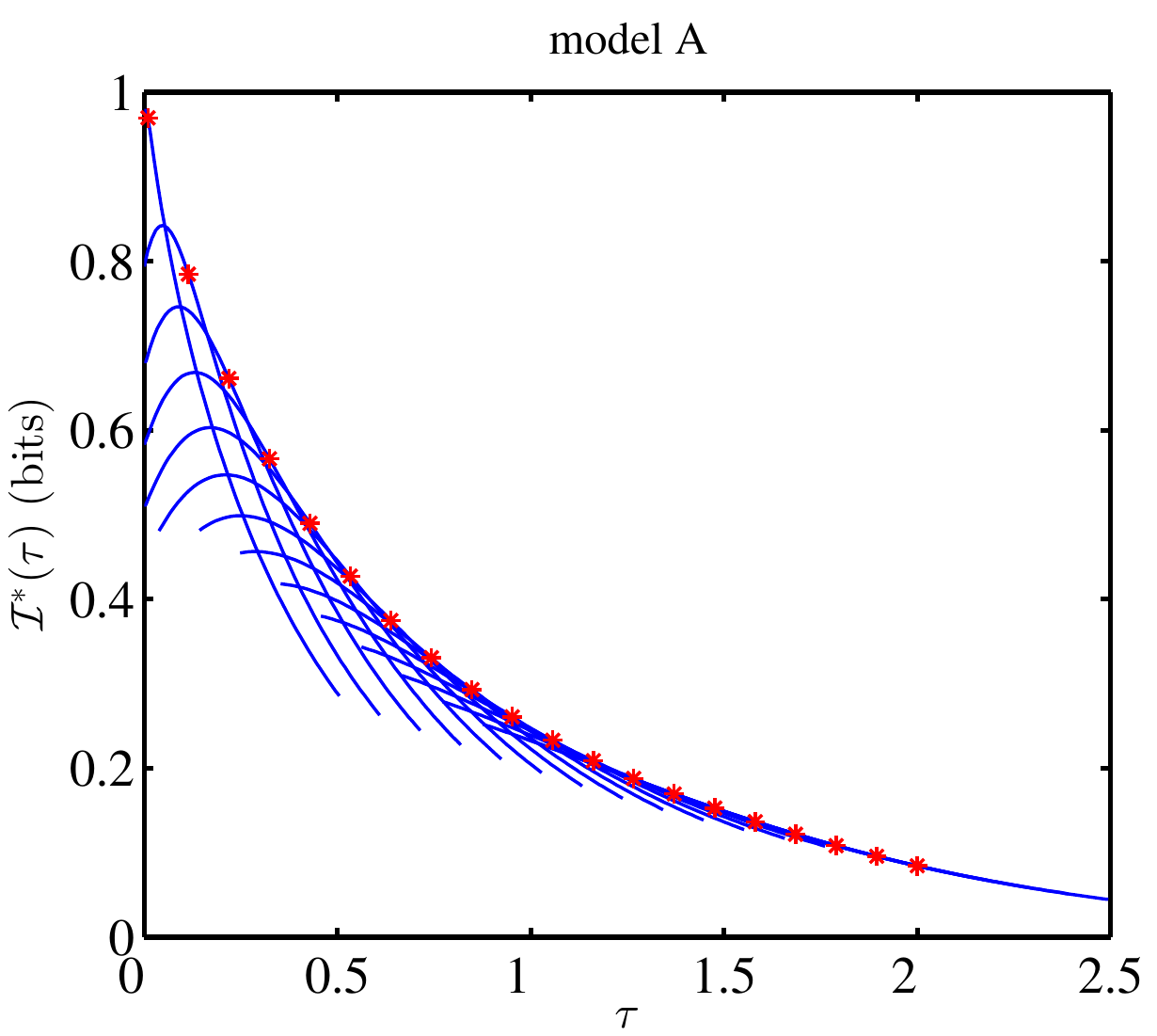}
\caption{An explicit construction of the optimal information curves presented in the \MS, here shown for model \mA.
The maximum value $\mathcal{I}^*(\tau)$ for each $\tau$ (red $*$) is obtained by optimizing the rates $u$ and $r$ at each $\tau$. The optimal rates can be different for each $\tau$. The blue continuous curves show the whole range of $\mathcal{I}(\tau)$: each of them intersects a $(\tau_{\mathcal{I}^*},\mathcal{I}^*(\tau))$ and is computed by using the corresponding optimal set of rates.} 
\label{hullfig}
\end{figure}

The results are obtained as hull plots, as presented in Fig.~\ref{hullfig} for model \mA.

\section{Derivation of mutual information for 2-bit, symmetric systems}
\label{appC}

In this Appendix we derive useful relations for the  mutual information in the case of $1$-bit symmetric systems. Consider any arbitrary distribution $p(a,b)$ where $\{a,b\}\in\{-1,+1\}$ and
the distribution enjoys symmetry under flipping $-1\leftrightarrow +1$.
In this case $p(+,+){=}p(-,-)$ and $p(+,-){=}p(-,+)$. For any such distribution
the mutual information greatly simplifies, as exploited in the 
analytic results associated with model \mA.

Let us define $-1<\eta<1$ such that
\beq
p(+,+)=p(-,-)=(1+\eta)/4,\\
p(-,+)=p(+,-)=(1-\eta)/4.
\eeq
From these, we see that
\beq
p(a)=p(b)=1/2
\eeq
and
\beq
I[a,b]&=&\sum_{a,b}p(a,b)\log_2 p(a,b)/(p(a)p(b))\notag\\
&=&\sum_{a,b}p(a,b)\log_2 4p(a,b)\notag\\
&=&2p(-,-)\log_2 4p(-,-)+2p(-,+)\log_2 4p(-,+)\notag\\
&=&\frac{1}{2}(1+\eta)\log_2 (1+\eta)+\frac{1}{2}(1-\eta)\log_2(1-\eta)
\eeq
as in the \MS, where we replace $\eta$ by the appropriate expression for $\mu$ in each model.

Note also that
\beq
p(a= b,b)-p(a\ne b,b)=\notag\\ 
p(+,+)-p(-,+)=\notag\\ 
1/4(1+\eta-1+\eta)=\eta/2.
\eeq

\section{Explicit calculation for model \mA}
\label{appD}

Model \mA describes a system in which $z$  up-regulates $x$ and $x$ is slaved to $z$. In this case we can diagonalize $\cal{L}$ analytically and calculate the mutual information. We set 
\begin{itemize}
\item $u_m{=}u_p{=}d_m{=}d_p{\equiv}u$,
\item $r_m{=}r_p{\equiv}r{=}1$,
\item $s_m{=}s_p{=}0$,
\end{itemize}
and the initial probability $p_0(y)$ is set equal to the steady state $p_{\infty}(y)$.

In this case the transition rate matrix $\mathcal{L}$ is given by
$$
\mathcal{L} = \left( \begin{array}{cccc}
                      u & -u & -r & 0 \\
		      -u & u+r & 0 & 0 \\
		      0 & 0 & u+r & -u \\
		      0 & -r & -u & u
                     \end{array}
	      \right)
$$
and its spectrum is 
\begin{equation}
 \label{eq:spectrum}
\{\lambda_1=0,\, \lambda_2=r,\, \lambda_3=2u,\, \lambda_4=r+2u\}.
\end{equation}
The largest relaxation rate of the system is given by the inverse of the first non-zero eigenvalue, which switches from $\lrelax{=}\lambda_3$ ($2u^*$) for small $\tau$ to $\lrelax{=}\lambda_2{=}\lambda_3$ ($2u^*{=}r^*$) for large $\tau$. 
This change in the rates that govern the relaxation times marks the crossover shown in Fig.~\ref{Fig2} in the \MS. \\
The left eigenvectors are
\begin{equation}
\label{eq:leigenvec}
\begin{cases}
u_1^{T} = (1,1,1,1), \\
u_2^{T}= (-1,\frac{-u+r}{u},\frac{+u-r}{u},1), \\
u_3^{T} = (-1,1,-1,1),\\
u_4^{T} = (1,\frac{-u-r}{u},\frac{-u-r}{u},1).
\end{cases}
\end{equation}
\begin{widetext}
The right eigenvectors are
\begin{equation}
\label{eq:reigenvec}
 \textstyle  v_1 =\frac{1}{2(r+2u)}\left(\begin{array}{c} r+u \\ +u  \\ +u  \\ r+u  \end{array}\right),
v_2 = \frac{1}{2(r-2u)}\left(\begin{array}{c} +u \\ +u  \\ -u  \\ -u  \end{array}\right),
v_3 =\frac{1}{2(r-2u)}\left(\begin{array}{c} -r+u \\ -u  \\ +u  \\ +r-u  \end{array}\right),
v_4 = \frac{1}{2(r+2u)}\left(\begin{array}{c} +u \\ -u  \\ -u  \\ +u  \end{array}\right).
\end{equation}
Using the expressions introduced in Appendix \ref{appC}, we find that $p(x',z)$ is given by the following $2\times2$ matrix:
\begin{equation}
p(x',z)= \left(
\displaystyle{\begin{array}{cc}
 \frac{\left(1+e^{-2 t u}\right) r^2+2 \left(-2 e^{-r t}+e^{-2ut}\right) r u-4 u^2}{4(r-2u)(r+2u)} & 
 \frac{\left(1-e^{-2ut}\right) r^2+2 \left(+2 e^{-r t}-e^{-2ut}\right) r u-4 u^2}{4(r-2u)(r+2u)} \\
\frac{\left(1-e^{-2ut}\right) r^2+2 \left(+2 e^{-r t}-e^{-2ut}\right) r u-4 u^2}{4(r-2u)(r+2u)} &
 \frac{\left(1+e^{-2ut}\right) r^2+2 \left(-2 e^{-r t}+e^{-2ut}\right) r u-4 u^2}{4(r-2u)(r+2u)}
\end{array}}
\right).
\end{equation}
We can then compute $I[x_t,z_0]$, which, after some algebraic manipulation, reads:
\begin{align}
I[x_t,z_0]&=\frac{1}{2}\left(1+\frac{-4 e^{-r t} r u+e^{-2 t u} r (r+2 u)}{(r-2 u) (r+2 u)}\right) \log_2\left[1+\frac{-4 e^{-r t} r u+e^{-2 t u} r (r+2 u)}{(r-2 u) (r+2 u)}\right]+ \notag\\
&+ \frac{1}{2}\left(1-\frac{-4 e^{-r t} r u+e^{-2 t u} r (r+2 u)}{(r-2 u) (r+2 u)}\right) \log_2\left[1-\frac{-4 e^{-r t} r u+e^{-2 t u} r (r+2 u)}{(r-2 u) (r+2 u)}\right].
\label{infomodelA}
\end{align}
\end{widetext}
If we introduce the quantity 
\beq
\mu=\frac{-4 e^{-r t} r u+e^{-2 t u} r (r+2 u)}{(r-2 u) (r+2 u)},
\eeq 
we recover Eq.~\ref{infoent} in the \MS. We recall that for large $\tau$ the optimal rates are $r^*{=}2u^*$: for this special case the expressions above may be evaluated
by Taylor expanding about $r{=}2u$ to find
\beq
\mu=e^{-rt}/2(1-2rt).
\eeq
We see that for $t{=}0$, $I[x_t,z_0]{=}1$ bit and, for long times, $\mu{\rightarrow}0$ and $I[x_t,z_0]{\rightarrow}0$. \\
Moreover, on all time-scales we find that in the stationary state $p(x'{=}z)/p(x'{=}-z){=}(r/u+1)>1$. 

From Eq.~\ref{infomodelA}, taking 
$ {{\delta I}\over {\delta t}}|_{t^*}{=}0 $
 and using the appropriate expression for $\mu$, we are able to find the optimal time lag for model \mA as 
 \beq
 t^*={\log\left(\frac{2u+r}{2r} \right)}/{(2 u - r)} {\rm \  for \ }r \neq2u,
 \eeq 
 or 
 $t^*=\displaystyle{\frac{1}{2r}}$, for $r=2u$,
 as presented in the \MS.

\section{Extension to model \mC}
\label{appE}
In model \mC we allow for all the eight rates to be nonzero and different from each other. Due to the symmetries of the system (e.g., relabeling the nodes and their rates), there are many parameter settings which result in equal mutual information. Qualitatively, these topologies may all be described as either
\begin{enumerate}
\item  $z$ activates $x$, which in turn represses $z$, or
\item  $z$ represses $x$, which in turn activates $z$.
\end{enumerate}
As an example, we consider a topology of the first type. { Numerical optimization allows us to observe that certain rates are zero. We exploit this observation to perform further analytical calculation. In the case of the topology of the first type, the rates that incorporate the numerical facts are the following:}
\begin{itemize}
\item $u_m=d_p$,
\item $u_p=d_m=0$,
\item $r_m=r_p$,
\item $s_p=s_m=0$.
\end{itemize}
Now the transition rate matrix $\mathcal{L}$ is given by
$$
\mathcal{L} = \left( \begin{array}{cccc}
                      u_m & 0 & -r_m & 0 \\
		      -u_m & r_m & 0 & 0 \\
		      0 & 0 & r_m & -u_m \\
		      0 & -r_m & 0 & u_m
                     \end{array}
	      \right)
$$
and its spectrum is 
\begin{equation}
\begin{cases}
 \label{eq:spectrum}
\lambda_1=0,\\ \lambda_2=r_m+u_m,\\ \lambda_3=\frac{1}{2} \left(r_m+u_m-\sqrt{r_m^2-6 r_m u_m+u_m^2}\right),\\ \lambda_4=\frac{1}{2} \left(r_m+u_m+\sqrt{r_m^2-6 r_m u_m+u_m^2}\right).
\end{cases}
\end{equation}

\begin{widetext}
For small $\tau$, the largest relaxation time is given by $1/\lambda_3$. For large $\tau$, $u_m^* = (3-2\sqrt{2}) r_m^*$ (where $r_m^*=1$), therefore $\lambda_3=\lambda_4$.

The left eigenvectors are
\begin{equation}
\label{eq:leigenvec}
\begin{cases}
u_1^{T} = (1,1,1,1), \\
u_2^{T} = (1,-\frac{r_m}{u_m},-\frac{r_m}{u_m},1), \\
u_3^{T} = (-1, \frac{+r_m-u_m-\sqrt{r_m^2-6 r_m u_m+u_m^2}}{2 r_m}, \frac{-r_m+u_m+\sqrt{r_m^2-6 r_m u_m+u_m^2}}{2 r_m}, 1),\\
u_4^{T} = (-1, \frac{+r_m-u_m+\sqrt{r_m^2-6 r_m u_m+u_m^2}}{2 r_m}, \frac{-r_m+u_m-\sqrt{r_m^2-6 r_m u_m+u_m^2}}{2 r_m}, 1).
\end{cases}
\end{equation}
The right eigenvectors are
\begin{equation*}
 \textstyle  v_1 =\frac{1}{2(r_m+u_m)}\left(\begin{array}{c} r_m \\ u_m \\ u_m \\ r_m \end{array}\right),
v_2 = \frac{1}{2(r_m+u_m)}\left(\begin{array}{c} +u_m \\ -u_m \\ -u_m \\ +u_m \end{array}\right), 
\end{equation*}
\begin{equation}
v_3 =\left(\begin{array}{c} \frac{-r_m+u_m-\sqrt{r_m^2-6 r_m u_m+u_m^2}}{4 \sqrt{r_m^2-6 r_m u_m+u_m^2}} \\ 
\frac{-u_m}{2 \sqrt{r_m^2-6 r_m u_m+u_m^2}} \\ 
\frac{+u_m}{2 \sqrt{r_m^2-6 r_m u_m+u_m^2}}  \\ 
\frac{+r_m-u_m+\sqrt{r_m^2-6 r_m u_m+u_m^2}}{4 \sqrt{r_m^2-6 r_m u_m+u_m^2}}  \end{array}\right),
v_4 = \left(\begin{array}{c} \frac{+r_m-u_m-\sqrt{r_m^2-6 r_m u_m+u_m^2}}{4 \sqrt{r_m^2-6 r_m u_m+u_m^2}}  \\ 
\frac{+u_m}{2 \sqrt{r_m^2-6 r_m u_m+u_m^2}} \\
 \frac{-u_m}{2 \sqrt{r_m^2-6 r_m u_m+u_m^2}}  \\ 
  \frac{-r_m+u_m+\sqrt{r_m^2-6 r_m u_m+u_m^2}}{4 \sqrt{r_m^2-6 r_m u_m+u_m^2}}  \end{array}\right).
\end{equation}
The mutual information is given by Eq.~\ref{infoent} in the \MS, with $\mu$ given by
\beq
\mu=\cosh({t \sqrt{u_m^2+r_m^2-6r_mu_m}})\frac{r_m-u_m}{r_m+u_m}+\sinh({t \sqrt{u_m^2+r_m^2-6r_mu_m}})\frac{u_m^2-r_m^2-4r_mu_m}{(r_m+u_m)\sqrt{u_m^2+r_m^2-6r_mu_m}}.
\eeq
\end{widetext}

For $t{=}0$,
$
\displaystyle{\mu_0{=}\frac{r_m-u_m}{r_m+u_m}}
$
 and $I[x_t,z_0]{=}1$ bit, as can be seen from Fig.~\ref{Fig2} \mC in the \MS. 
For large $t$ we can define 
 \beq
 x=t \sqrt{u_m^2+r_m^2-6r_mu_m}
 \eeq
and, noting that 
\beq
\lim_{x^* \rightarrow 0}{\sinh{x^*}\over{x^*}}=1,
\eeq
we obtain the optimal 
 \beq
 \mu^*=\frac{r_m-u_m}{r_m+u_m}+\frac{u_m^2-r_m^2-4r_mu_m}{r_m+u_m}t.
 \eeq

\section{Model \tA}
\label{appF}

Model \tA is the same as model \mA, except for the fact that the initial probability $p_0(y)$
is not given by the steady state but is optimized. 
In order to optimize $I[x_t,z_0]$, one demands that the entropy $S[p(z)]$ is equal to $1$ 
\footnote{For any distribution $p(a,b)$, $I[a,b]<\min\{S[a],S[b]\}$. In this case, if we want $I[x_t,z_0]$ to reach $1$ bit, it must be the case that $S[x_t]$ and $S[z_0]$ are both $1$ bit.}. This, together with the symmetry in $-1\leftrightarrow +1$ for $x$ and $z$, constrains the form of the \prior to be parameterized as
\begin{equation}
p_0(x,z) = \left(\frac{1+\mu_0 }{4},\frac{1-\mu_0 }{4},\frac{1-\mu_0 }{4},\frac{1+\mu_0 }{4}\right).
\end{equation}
\begin{widetext}
The probability $p(x',z)$ then reads
\begin{equation}
p(x',z)=\left(
\begin{array}{cc}
 \frac{r+e^{-2 t u} r-2 u+e^{-r t} (-r+ r \mu_0 -2 u \mu_0 )}{4 (r-2 u)} & 
 \frac{r-e^{-2 t u} r-2 u+e^{-r t} (+ r-r \mu_0 +2 u \mu_0 )}{4 (r-2 u)} \\
 \frac{r-e^{-2 t u} r-2 u+e^{-r t} (+ r-r \mu_0 +2 u \mu_0 )}{4 (r-2 u)} &
  \frac{r+e^{-2 t u} r-2 u+e^{-r t} (-r + r\mu_0 -2 u \mu_0 )}{4 (r-2 u)}
\end{array}
\right),
\end{equation}
and we can explicitly compute the mutual information: 
\begin{align}
I[x_t,z_0]&=\frac{1}{2}\left(1+ \mu_0e^{-rt} + \frac{r}{r-2u} (e^{-2u t}-e^{-rt})\right) \log_2\left[1+ \mu_0 e^{-rt} + \frac{r}{r-2u} (e^{-2u t}-e^{-rt})\right]+ \notag\\
&+ \frac{1}{2}\left(1-\mu_0 e^{-rt} + \frac{r}{r-2u} (e^{-2u t}-e^{-rt}) \right) \log_2\left[1-\mu_0 e^{-rt} + \frac{r}{r-2u} (e^{-2u t}-e^{-rt})\right].
\end{align}
\end{widetext}

The above equation can again be rewritten as in Eq.~\ref{infoent} if we introduce 
\beq
\mu=\mu_0 e^{-rt} + \frac{r}{r-2u} (e^{-2u t}-e^{-rt}).
\eeq
 For $t{=}0$, $\mu{=}\mu_0$ and the information is maximized by $\mu{=}\mu_0{=}\pm 1$. 

\section{Models \mB and \tB}
\label{appG}

\begin{figure}[!ht]
\includegraphics[scale=0.7]{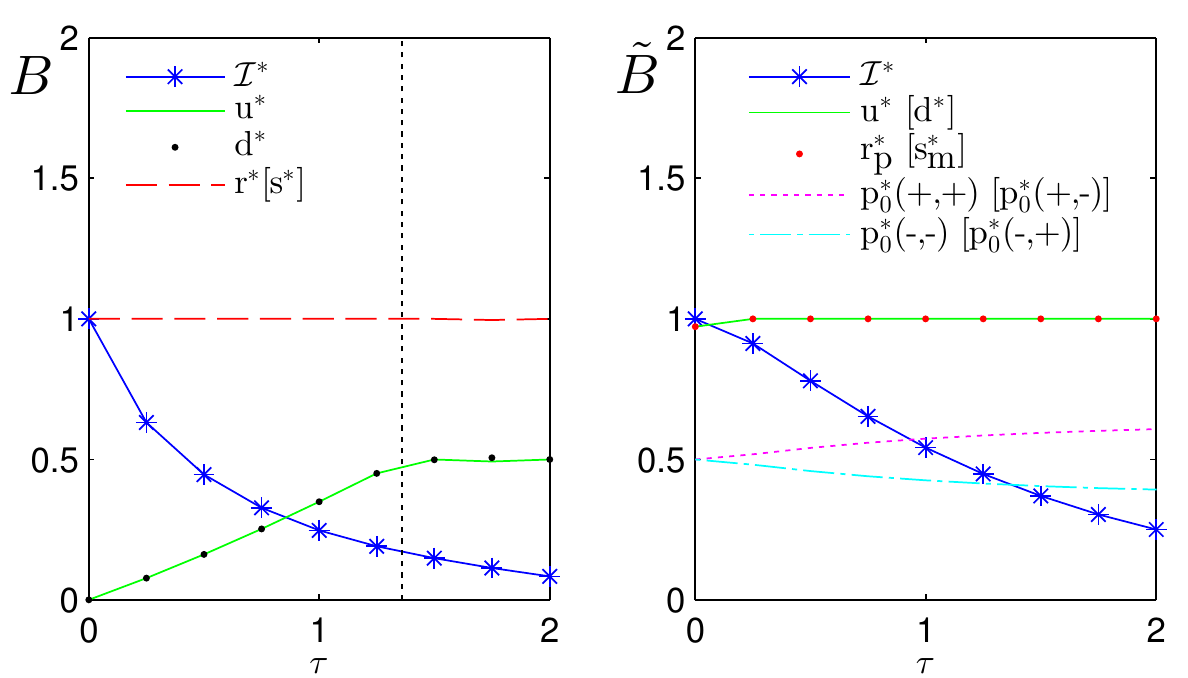}
\caption{\label{modelB} Example of optimal parameters, along with transmitted information  $\mathcal{I}^*$, for models \mB and \tB. Parameters in square brackets refer to alternative optimal topologies where $z$ down-regulates $x$ instead of up-regulating it (see Fig.~\ref{FigPan_SI} \mB and \tB). Subscripts $m,p$ are omitted when $x_m{=}x_p$ (with $x{=}u,d,r,s$). 
In model \tB the optimal rates  do not change with $\tau$ and the optimal \prior is split slightly unevenly between
the beginning and end state. }
\end{figure}

\begin{figure}[!ht]
\includegraphics[scale=0.7]{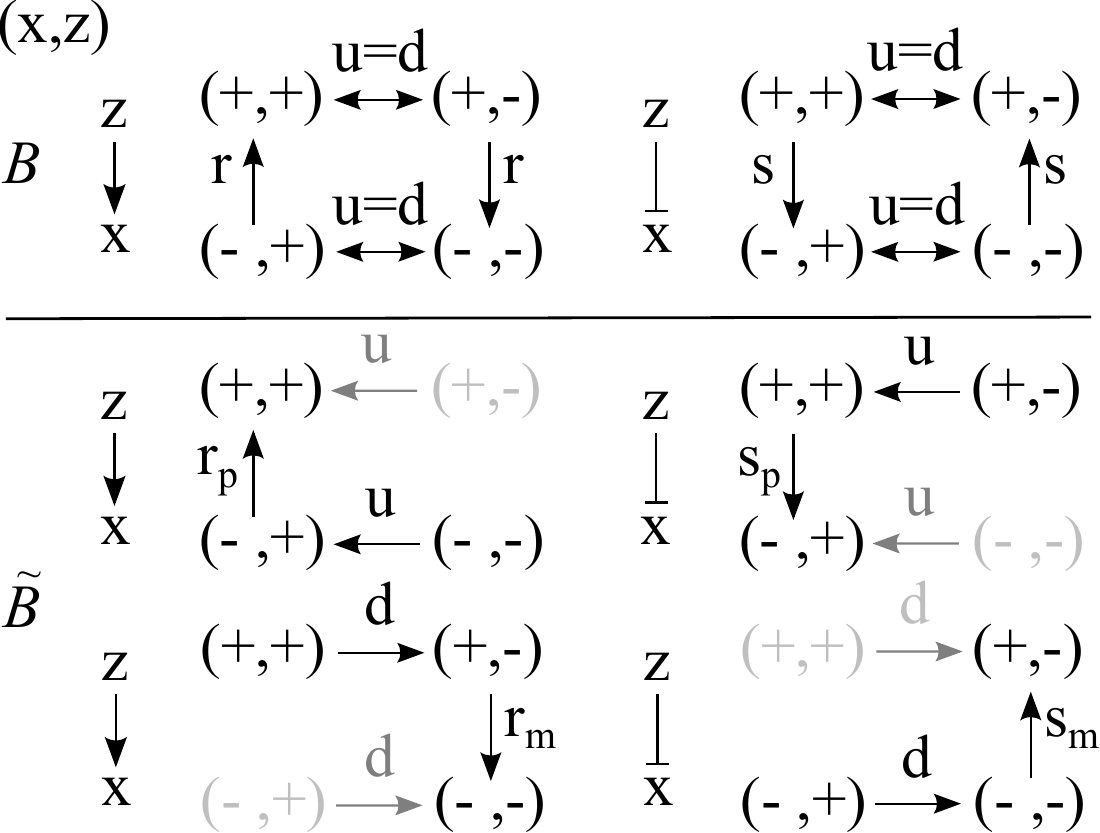}
\caption{\label{FigPan_SI} Full set of optimal topologies and dynamics for models \mB and \tB. The topologies of these optimal networks were found by inspecting the optimal rates manually. In model \tB some states (shown in gray in the figure) are never visited. }
\end{figure}

Models \mB  and  \tB are extensions of model \mA and  \tA, respectively, where we allow $z$ to also be a repressor of $x$ 
($s_m\neq0, s_p\neq 0$, and we no longer demand $r_p=r_m$). As in models \mA and \tA, we do not allow  feedback from $x$ to $z$, meaning that the transition rates for $z$ do not depend on the state of $x$ (i.e. $u_m{=}u_p{\equiv}u$ and $d_p{=}d_m{\equiv}d$). The results for the optimal parameters and topologies of these models are very similar to models \mA and \tA and are not shown in the \MS. We plot the corresponding results in Fig.~\ref{modelB}. We note that in model \tB we have absorbing states, similar to those of model \tC shown in the \MS. As in model \tC, the symmetries of the problem allow many equivalent (under permutation of labels) parameter settings: we show all of them in Fig.~\ref{FigPan_SI}. However, unlike in model \tC where all states are visited, in each optimal setting of model \tB one state (gray in Fig.~\ref{FigPan_SI}) is never visited.

\vspace{- 0.5 cm}
\section{Model \tC}
\label{appH}

\begin{figure}[!ht]
\includegraphics[scale=0.7]{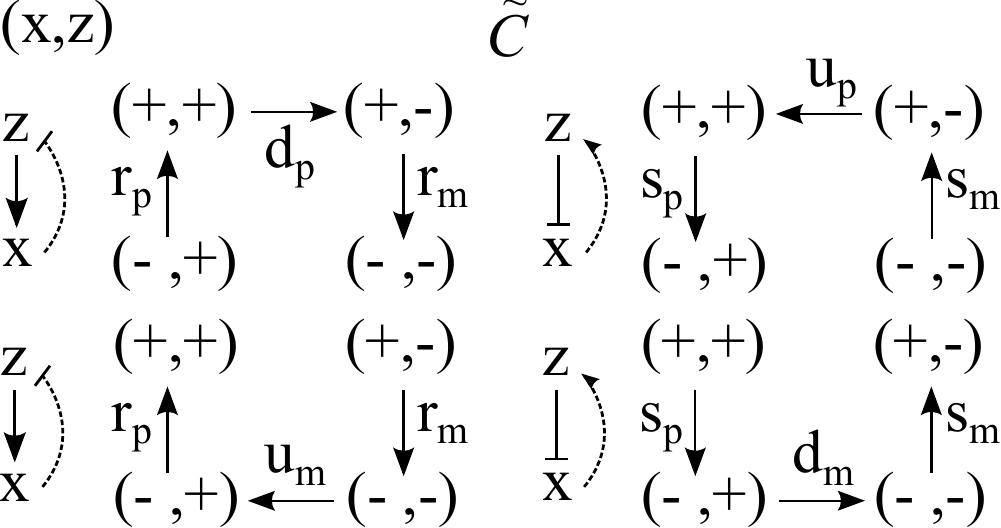}
\caption{\label{FigPan_SI_Ctilde} Optimal topologies and dynamics for model \tC, where the steady-state assumption is relaxed and feedback is present. The topologies of these optimal networks were found by inspecting the optimal rates manually. The optimal \prior is split slightly unevenly between
the beginning and end state.
}
\end{figure}

Model \tC is the same as model \mC, except for the fact that the steady-state assumption is now relaxed and the system is optimized also over the \prior (as in models \tA and \tB). 
The results are discussed in detail in the \MS; here, we simply show in Fig.~\ref{FigPan_SI_Ctilde} the four maximally informative topologies that have been found: each of them can be labeled as a ``push-pull'' network and features an absorbing state. 

\newpage

\bibliography{timedepinfo_biblio}

\end{document}